\documentclass{article}

\usepackage{arxiv}

\usepackage[utf8]{inputenc} 
\usepackage[T1]{fontenc}    
\usepackage{url}            
\usepackage{booktabs}       
\usepackage{amsfonts}       
\usepackage{nicefrac}       
\usepackage{microtype}      
\usepackage{lipsum}
\usepackage{graphicx}
\usepackage{textcomp}
\graphicspath{ {./figs/} }

\usepackage{fancyhdr}
\usepackage{hyperref}
\hypersetup{
    colorlinks=true,
    linkcolor=blue,
    citecolor=blue,
    filecolor=magenta,      
    urlcolor=cyan,
    pdftitle={Overleaf Example},
    pdfpagemode=FullScreen,
}

\title{Continuous risk assessment in secure DevOps}

\author{
 Ricardo M. Czekster \\
  School of Computer Science and Digital Technologies\\
  Aston University\\
  Birmingham, United Kingdom, B4 7ET \\
  \texttt{r.meloczekster@aston.ac.uk} \\
}

\begin{document}
\maketitle
\begin{abstract}
DevOps (development and operations), has significantly changed the way to overcome deficiencies for delivering high-quality software to production environments.
Past years witnessed an increased interest in embedding DevOps with cybersecurity in an approach dubbed secure DevOps.
However, as the practices and guidance mature, teams must consider them within a broader risk context.
We argue here how secure DevOps could profit from engaging with risk related activities within organisations.
We focus on combining Risk Assessment (RA), particularly Threat Modelling (TM) and apply security considerations early in the software life-cycle.
Our contribution provides a roadmap for enacting secure DevOps alongside risk objectives, devising informed ways to improve TM and establishing effective security underpinnings in organisations focusing on software products and services.
We aim to outline proven methods over the literature on the subject discussing case studies, technologies, and tools.
It presents a case study for a real-world inspired organisation employing the proposed approach with a discussion.
Enforcing these novel mechanisms centred on security requires investment, training, and stakeholder engagement.
It requires understanding the actual benefits of automation in light of Continuous Integration/Continuous Delivery settings that improve the overall quality of software solutions reaching the market.
\end{abstract}


\section{Introduction}
Modern Software Development Life-Cycle (SDLC) evolved from  strict Waterfall methodology or specializations like the V-model (that focused on testing for each phase) to modern Agile approaches namely eXtreme Programming (XP), Scrum, Kanban~\cite{theocharis2015water,saleh2017comparative}, and hybridisations~\cite{stoica2016analyzing,kuhrmann2017hybrid} where organisations `cherry pick' practices they believe contributes to productivity.
Stakeholders must ensure adequate end-product level-of-service reaching customers through stringent observance to quality properties that prevent serious defects or protects users from malicious behaviours.
Examples are performance, usability/intuitiveness, energy efficiency, and other ``-ities/-ilities'', ie, availability, reliability, integrity, maintainability, and agility, in a non-exhaustive list~\cite{bass2021software,glinz2007non,chung2012non}.
We focus here on another important quality property given by security.
This concern stands out as a crucial ``non-requirement''~\cite{hohpe2020software}, ie, everyone simply assumes that a system or service are thought to be automatically safe and secure.
Security (used interchangeably here with cybersecurity)\footnote{This work entails offering protections to users at any level, ie, cyber-physical, encompassing any malicious attempt not anticipated by requirements.} entails observing so-called CIA triad: Confidentiality, Integrity, and Availability~\cite{avizienis2004basic}.

Developers do not wish to sit down and read risk related documentation on a higher abstraction level (ie, management, governance, leadership).
They want to start thinking about the solution straight away, usually by means of Threat modelling (TM) systems and services, using what they know, eg, Data Flow Diagrams (DFD), Attack/Threat Trees, Unified modelling Language (UML) diagrams (Interaction Diagrams, and a host of other diagrams), and so on.

A modern approach of SDLC is called DevOps~\cite{fitzgerald2017continuous,leite2019survey,kim2021devops}, the idea of crafting solutions that considers development to deployment (operations).
The four agreed principles are Culture, Automation, Measurement, and Sharing (CAMS)~\cite{humble2011enterprises}.
One distinct variant is secure DevOps that aims to seamlessly embed security into the SDLC through early TM and Shift-Left Security~\cite{battina2017best,mansfield2018devops,rajapakse2022challenges}, also called Start-Left~\cite{tarandach2020threat}\footnote{Note that these \textit{buzzwords} have been abused by vendors, thus significantly losing its core meaning, which is to embed security principles early on.}.
Secure DevOps undoubtedly poses challenges for adoption in software teams as discussed by \cite{myrbakken2017devsecops} and \cite{larrucea2019dealing}.

Risk Management (RM) and Risk Assessment (RA) cannot be detached from these considerations as they belong to a broader risk-based approach within organisations that allocate resources to combat potential threats, remediate cyber-attacks, and make sure security controls are working effectively, among other tasks.
We argue here, however, the need for understanding the context and motivations on how RM/
RA intertwines with TM.

\begin{figure}[!ht]
\centering
\includegraphics[width=.9\linewidth]{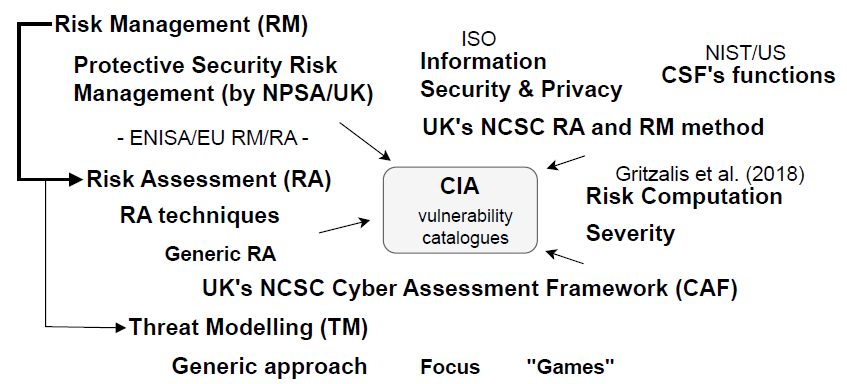}
\caption{Overview of approaches and interplay among CIA, RM, RA, and TM.}
\label{fig:approaches-overview}
\end{figure}

We provide an overview of the general process for understanding the interplay among RM, RA and TM in Figure~\ref{fig:approaches-overview}.
It outlines how the approaches are inter-related and the geographically distributed institutions behind the initiatives (where leaders are currently the US, the UK, and the European Union).
TM can be seen as a RA technique within a broader RM context.
It has gained significant traction in software development community as it provides means and tools to understands most-likely threats to systems and services.

As motivation, we have identified inherent difficulties of working with domain experts across fields when enhancing the cybersecurity posture of software-based solutions.
For instance, one setting may have risk analysts (broader risk panorama in an organisational level), programmers/testers, software architects, governance, compliance and security officers (some with coding skills, others only with security/privacy domain expertise), and so on, in heterogeneous settings in terms of skill-sets and capabilities.
We outline next our contributions:

\begin{itemize}
   \item Thorough discussion highlighting how secure DevOps practitioners should incorporate Shift-Left Security practices in a broader risk context and how these elements are intermixed altogether.
   \item Review on how RM/RA approaches mixes with secure DevOps and how developers and risk analysts could combine efforts while embracing Continuous Integration/Continuous Delivery-Deployment (CI/CD).
   \item Showcase the importance of considering risk management, risk assessment and threat modelling approaches in software development life-cycle as proposed in secure DevOps contexts.
   \item Perspective on joint team/\,management/\,customer about the potential overhead imposed by new features entering the Product Backlog and the inherent issues to take into account in terms of effort and feasibility aligned with budgetary constraints.
   \item Comprehensive examination of beneficial practices to enhance the overall software quality requirements with focus on cybersecurity of features reaching production environment within CI/CD mechanisms and Application Security Testing (AST) tools and methods.
\end{itemize}

To the extent of the literature survey presented here, we were unable to find work outlining the desired cross-fertilisation of RM/RA with TM in secure DevOps contexts with effective practices and recommendations.
One work aligned with this one was proposed by Dupont et al. (2022)~\cite{dupont2022product}, where they provided a discussion of mixing RA with secure DevOps, however, the approach discussed here goes further in providing context for RM/RA altogether.
Another close aligned work was discussed by Zografopoulos et al. (2021)~\cite{zografopoulos2021cyber}, where they mixed RA and TM in a cyber-physical energy systems context.
The sheer amount of documentation, best practices, compliance, recommendations, guidance on TM/RA, etc., hinders productivity and slows feature development/deployment into production environments.

We stress that this paper does not aim to be an authoritative cybersecurity ``must-follow'' account of how to best employ RM, RA, and TM within organisations.
It serves as a compilation of related documents or as guidance to provoke further in-depth risk related investigations and matching approaches to organisational objectives.
As it is the case for many security publications, there is no `silver bullet', `panacea', enforced checklists\footnote{The Open Worldwide Application Security Project (OWASP) does provide a broad and interesting list of so-called \textit{cheat sheets}: \url{https://cheatsheetseries.owasp.org/}, with useful steps to follow.}, or general rule-set to blindly follow and expect systems and services to be automatically secure and protected from threat agents.

This work is organised as follows.
Section~\ref{s:context} will set the context and Section~\ref{s:eff} will discuss how to incorporate effective TM/RA in secure DevOps.
In Section~\ref{s:recommendations} we present our recommendations and guidelines with practical implications.
Finally, Section~\ref{s:conc} delineates our conclusion, a roadmap for adopting secure DevOps altogether, and future work.

\section{Context: tackling risks}\label{s:context}
Basic security principles point out to practitioners that any attempt of abusing a system or service are related to the so-called CIA triad: Confidentiality, Integrity, and Availability.
Modern authors~\cite{stallings2018effective} include Authenticity, Accountability, Privacy, Trustworthiness, and Non-Repudiation, to factor in requirements for auditing, forensics, and uniquely identifying individuals.

The sheer amount of risk and security related guidance, recommendations, and standards for risk management (RM) and risk assessment (RA) is overwhelming, especially for inexperienced stakeholders.
Focusing on the most significant ones, there is ISO 31000:2018 tackling risk management (and accompanying ISO/IEC 31010:2019 -- risk assessment techniques), ISO 27005:2022 for information security risk management, the European Network and Information Security Agency (ENISA) RM/RA and interoperability documents, NIST SP 800-37 for Risk Management Framework for Information Systems and organisations, and NIST SP 800-30~\cite{nist2012guide}, a Guide for Conducting Risk Assessments from the NIST Risk Management Framework (RMF), and the NIST SP 800-53, Security and Privacy Controls for Information Systems and organisations.

\begin{figure*}[!ht]
\centering
\includegraphics[width=.95\linewidth]{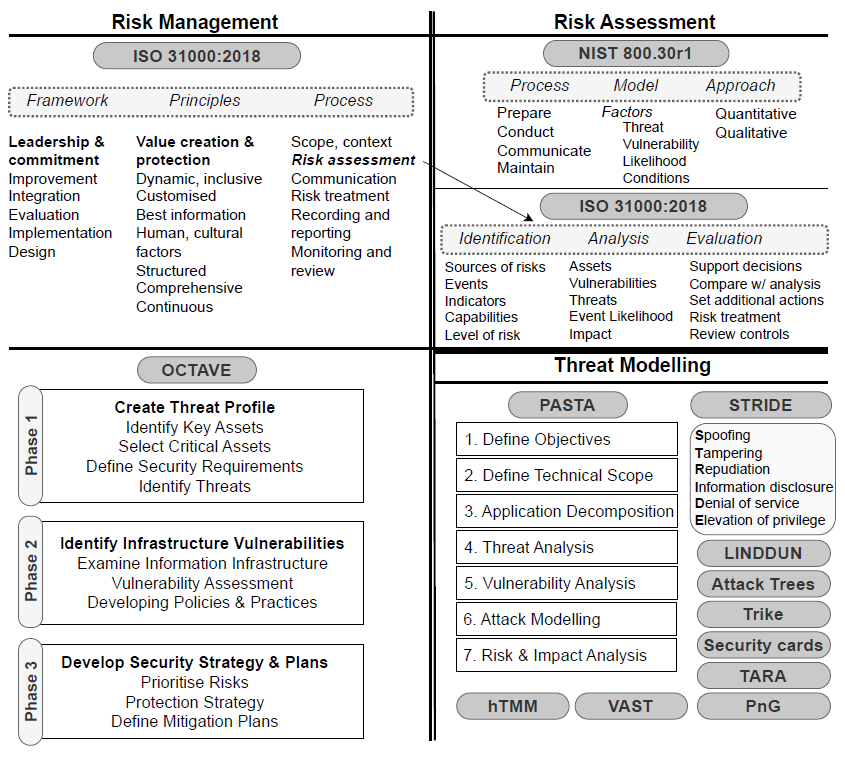}
\caption{Shared concepts and ideas of RM, RA, and TM with a summary of steps.}
\label{fig:risk-mgmt-risk-assmt}
\end{figure*}

RM and RA methodologies share concepts, for example, Figure~\ref{fig:risk-mgmt-risk-assmt} shows a juxtaposition among ISO 31000, OCTAVE~\cite{alberts2003managing}, and NIST 800.30r1 and the Process of Attack Simulation and Threat Analysis (PASTA)~\cite{ucedavelez2015risk}.
In ISO 31000:2018, risk is the effect (ie, deviation from expected) of \textit{uncertainty} on objectives.
It focuses on Framework (governance, leadership, commitment, improvement), Principles (value creation \& protection), and Process (context, risk communication, risk assessment -- risk identification, analysis, evaluation -- risk treatment, and reporting/monitoring risk in a continuous fashion.
The document defines risk sources, events, consequences, likelihood, impact, and controls.

For NIST, additionally, they have conceived the Cybersecurity Framework\footnote{Link: \url{https://www.nist.gov/cyberframework}.} (CSF 2.0 -- Feb/2024), a document that provides guidance for managing cybersecurity risks that could be integrated with previous NIST related RA methodologies.
The CSF core outlines functions such as Govern, Identify, Protect, Detect, Respond, and Recover.
As a noticeable difference from previous versions, it now recognises \textit{governance} as a dimension: \textit{``The GOVERN Function supports organisational risk communication with executives. Executives’ discussions involve strategy, particularly how cybersecurity-related uncertainties might affect the achievement of organisational objectives''.}

The UK's National Cyber Security Centre (NCSC) and the newly created (2024) National Protective Security Authority (NPSA) offer RM/RA recommendations.
NCSC offers guidance on RM that is inspired on ISO 27005 with the provision of a so-called RM Toolbox encompassing RM Information, RA Approaches (system-driven and component-driven), Assurance, and Tools \& Methods (which lists Attack Trees, TM, and Scenarios).
It offers a basic guidance on RA for people without experience in risk analysis and also on basic TM and Attack Tree modelling.
It has published the NCSC Cyber Assessment Framework (CAF) guidance as embracing UK's National Cyber Strategy, a new initiative for improving government cyber security.
NPSA has devised a document called Protective Security Risk Management, a two-page guidance outlining eight steps for conducting a broad RA on assets and information management systems.
For NPSA, \textit{``risks are identified threats or vulnerabilities, aligned to assets, that have been assessed for their likelihood (of the threat event occurring) and impact.''}

\begin{figure*}[!ht]
\centering
\includegraphics[width=1\linewidth]{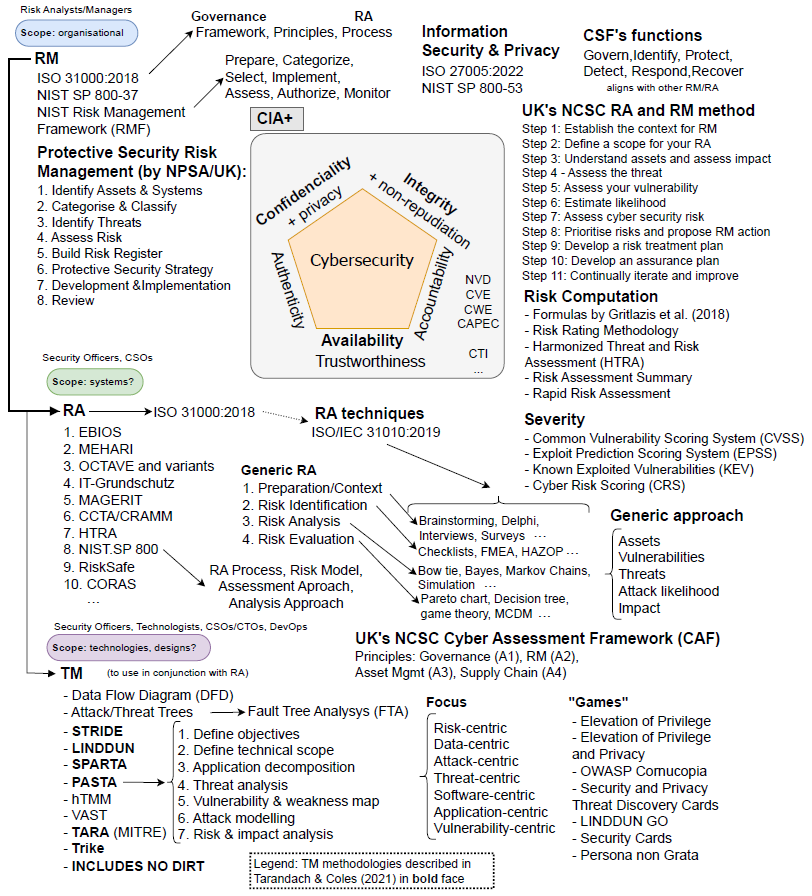}
\caption{Approaches and guidance documents and standards for RM, RA, and TM.}
\label{fig:approaches-all}
\end{figure*}

Figure~\ref{fig:approaches-all} conveys concepts and notions explained herein tangent to RM, RA, and TM, among other (it expands the key notions as presented in Figure~\ref{fig:approaches-overview}).

In terms of RA methodologies and risk computation formulas, \cite{gritzalis2018exiting} discuss classes and characteristics of selected approaches.
It focuses on the following RA methodologies: 
\begin{enumerate}
   \item \textit{Expression des Besoins et Identification des Objectifs de Sécurité} (EBIOS)
   \item MEthod for Harmonized Analysis of RIsk (MEHARI)
   \item Operationally Critical Threat and Vulnerability Evaluation (OCTAVE) and variants (OCTAVE Allegro, OCTAVE-S)
   \item IT-Grundschutz 
   \item \textit{Metodología de Análisis y Gestión de Riesgos de los Sistemas de Información} (MAGERIT)
   \item Central Computing and Telecommunications Agency Risk Analysis and Management Method (CRAMM)
   \item Harmonized Threat Risk Assessment (HTRA)
   \item NIST.SP 800
   \item RiskSafe
   \item CORAS
\end{enumerate}

It is worth mentioning that some methods are obsolete, as they have not been used in any modern risk assessment application domains in recent years, the case for example of CRAMM.
More recently, \cite{ekstedt2023yet} proposed a RA approach entitled Yet Another Cybersecurity Risk Assessment Framework (Yacraf).
Its objective is to help organisations with more decision support additionally offering a risk calculation formalisation.Another framework left out the analysis was the Factor Analysis of Information Risk (FAIR\texttrademark)~\cite{freund2014measuring,jones2006introduction}\footnote{Additionally, consult Open FAIR\texttrademark~ Risk Analysis Process Guide, Version 1.1 at \url{https://publications.opengroup.org/g180}, published by the Open Group.}, that aims to provide quantitative means for risk assessment tailored for information security.

Another interesting outcome in \cite{gritzalis2018exiting} is an attempt to amalgamate RA's phases altogether within four general items: 1. Preparation (ISO 31000:2018 calls this Context), 2. Risk Identification, 3. Risk Analysis, and 4. Risk Evaluation.
As a matter of fact, a myriad of proposed RA throughout the years presents these four phases, perhaps with minimal differences.

OWASP has proposed a Risk Rating Methodology for computing risk-based on guidance available in NIST 800-30, HTRA, and Mozilla's Risk Assessment Summary and Rapid Risk Assessment.
It is applicable to online applications requiring CIA considerations and uses the standard risk computation given by\footnote{For additional formulas and discussion, please consult \cite{gritzalis2018exiting}.}: $$Risk = Likelihood * Impact$$

This formula serves as the basis for many quantitative risk analysis in the literature.
A hybrid approach is Threat Analysis and Risk Assessment (TARA) widely used in the automotive industry alongside ISO 21434:2021 (Road vehicles -- Cybersecurity engineering)~\cite{marksteiner2023tara,dobaj2023towards}.
Not to confuse with Threat Assessment and Remediation Analysis (TARA) by MITRE Corporation\footnote{Link: \url{https://www.mitre.org/news-insights/publication/threat-assessment-and-remediation-analysis-tara}.} that specialises in the identification and ranking of attack vectors based on assessed risk~\cite{wynn2014threat}.
This approach has been used in the US under military, air force, and naval applications with success.
Intel has come up with an approach called Threat Agent Risk Assessment (TARA), however, there is inconsistent bibliography about it, so we will not focus on explaining in here.
RA has been applied to several contexts and applications, for instance, to industrial contexts~\cite{ralston2007cyber}, Smart Homes~\cite{manandhar2020towards} and Smart Buildings~\cite{mace2020smart,czekster2021cybersecurity}, critical infrastructure~\cite{pate2018cyber}, and energy systems~\cite{zografopoulos2021cyber}.

For addressing risk severity, a host of techniques are available, for instance, Common Vulnerability Scoring System (CVSS), EPSS (Exploit Prediction Scoring System), Known Exploited Vulnerabilities (KEV), and Cyber Risk Scoring (CRS).
The Software Engineering Institute (SEI) at Carnegie Mellon University/US has produced a White Paper outlining 12 available methods for TM\footnote{Link: \url{https://insights.sei.cmu.edu/blog/threat-modelling-12-available-methods/}.}: STRIDE, PASTA, LINDDUN, CVSS\footnote{Observe that CVSS is not considered a RA method, but a risk severity scoring system.}, Attack Trees, Persona non Grata, Security Cards, hTMM (Hybrid TM Method), Quantitative TMM, Trike, VAST (Visual, Agile, and Simple Threat) modelling, and OCTAVE\footnote{Note that LINDDUN focuses on privacy related threats whereas OCTAVE is a full-fledged and established RM/RA approach, and listing it as TM begs further discussion.}.
More recently, an addition to STRIDE was the STRIDE-LM (Lateral Movement) Threat Model (to account for LM being defined as a way of ``expanding control over the target network beyond the initial point of compromise''.)\footnote{Link: \url{https://csf.tools/reference/stride-lm/}.}

\begin{figure}[!ht]
\centering
\includegraphics[width=.8\linewidth]{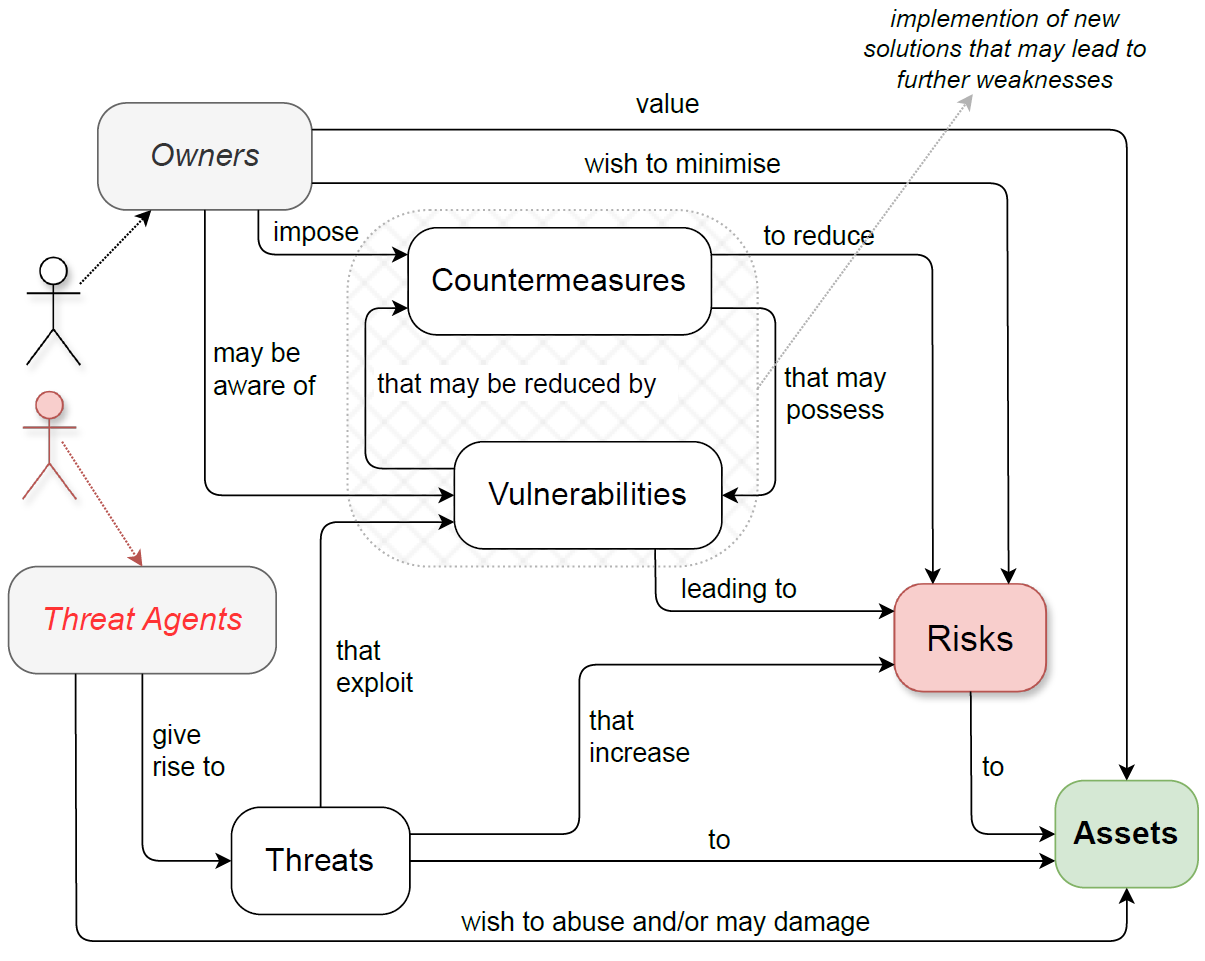}
\caption{Communicating risk related underpinnings with the Common Criteria -- adapted from \cite{ekstedt2011modelling}.}
\label{fig:common-criteria}
\end{figure}

The Common Criteria for Information Technology Security Evaluation (known simply as Common Criteria or CC) is an international standard (ISO/IEC 15408) for cybersecurity\footnote{Link: \url{https://www.commoncriteriaportal.org/index.cfm}.}.
Figure~\ref{fig:common-criteria} shows an adaptation of CC presenting major actors (Owners, Threat Agents), and relationship with Risks and Assets.

The figure helps understanding cybersecurity concepts by non-experts as it outlines the key relationships among Owners, Threat Agents, and Threats leading to Risks over Assets.
It also highlights how developers might introduce vulnerabilities into systems and services, ie, when patching and updating software, among other tasks. 

\subsection{Discussion}
One aspect shared by many of these methodologies, frameworks, and standards is that they attempt to be generally applicable to a host of situations requiring close attention to risks, threats, and vulnerabilities over assets.
It is a known fact that the security community has overgrown throughout the years and different ways of tackling risks have emerged, as seen by ISO/IEC, NIST/US, NCSC/UK, ENISA/EU and many other collaborations out of research groups, companies, and individuals.
To help our readers choosing methodologies and approaches for risk, we comment on the following guidance:
\begin{enumerate}
   \item If one is simply overwhelmed by the sheer available documentation: start with the CC, which is useful to understand the tasks for tackling risks over assets under threats in organisations. The work by Tarandach and Coles (2020)\cite{tarandach2020threat} also provides a simple view of security terminology aligned with the core ideas behind the CC (adding the concepts of system, data, value, and functionality).
   \begin{itemize}
       \item In an organisational level, one will need to select a proper RM methodology out of available options.
       \item The ISO 31000:2018 standard strives for conciseness: it has about 17 pages outlining risk and surrounding concepts in a simple to understand fashion.
   \end{itemize}
   \item For choosing a RA methodology, start by selecting the geographic region aligned to your objectives and budget. Options are 1. US (NIST); 2. UK (NCSC); 3. Europe (ENISA) -- as listed, there are approaches based on Spain, France, and Germany, among other. They have free and ready to use guidance -- mind that some RA methodologies have licensing and/or support costs (for instance, OCTAVE, IT-Grundschutz, MAGERIT, CRAMM, and RiskSave).
\end{enumerate}

\section{Intermixing RM, RA, and TM in secure DevOps}\label{s:eff}
Secure DevOps rapidly organised itself around DevOps as it became clear the need to insert security into software as early as possible.
The community has come up with the DevSecOps Manifesto\footnote{Link: \url{https://www.devsecops.org/}.} outlining practices and recommendations.
Similar approaches were also competing for attention, the case of SecDevOps~\cite{mohan2016secdevops}, secure DevOps~\cite{yasar2016integrate} (which is the terminology followed here), or even Rugged DevOps\footnote{Link: \url{http://ruggedsoftware.org/}.}.
The work by Lombardi and Fanton (2023)~\cite{lombardi2023devops} go one step beyond arguing that secure DevOps is not enough, teams require effective shift-left security practices in an approach they refer to as CyberDevOps.

There is a myriad of challenges for working with secure DevOps and TM as discussed by Jayakody and Wijayanayake (2021)~\cite{jayakody2021challenges} that outlined in a literature review: organisational culture changes to effectively absorb secure DevOps ideas, difficulties when finding experienced professionals, lack of management support, adopting the process, changing the mindset required for secure DevOps, issues for replicating complex technology environments, lack of collaboration, concerns when establishing a development culture, inherent complexities in software development and mismatch with secure DevOps, legacy systems, and increased project costs associated.
The work by \cite{valani2018rethinking} outlined issues to keep up project delivery velocity whilst maintaining TM and associated challenges for updating models to better scale it with automation and traceability features in secure DevOps.

Incorporating security into software engineering has been a concern discussed early in computing~\cite{devanbu2000software,mcgraw2004software,kern2007foundations}, with ramifications in system design practices and mechanisms on building systems ``the right way'' from the onset of projects~\cite{viega2001building}.
More recently, attention has been diverted into modern approaches, usually led by secure DevOps~\cite{rajapakse2022challenges} and incorporating it with TM practices and resources~\cite{myrbakken2017devsecops,xiong2019threat}.
The work by Battina (2017)~\cite{battina2017best} discussed best practices for incorporating security into DevOps, highlighting: 1. integrating with identity and access management and (secure) code review, 2. fitting it with governance, 3. effective vulnerability management, 4. automation, 5. validation, 6. network segmentation (more technical), and 7. use least privilege approaches.

The book by Shostack (2014)\cite{shostack2014threat} has thoroughly discussed TM in practical terms and how to design systems for security with applications to real-world systems.
One outcome was to discuss the overall TM process and propose four questions that each process must answer:
\begin{enumerate}
  \item What are we working on?
  \item What can go wrong?
  \item What are we going to do about it?
  \item Did we do a good enough job?
\end{enumerate}

This has become a key tenet of the TM Manifesto\footnote{Link: \url{https://www.threatmodellingmanifesto.org/}.}, among other substantial elements for establishing the necessary  context to sustain a TM-based process within organisations.
In a high-level perspective, note that the same questions could be asked by risk-based stakeholders without any loss; only the scope is broader.
Tarandach and Coles (2020)~\cite{tarandach2020threat} compiled a hands-on approach to TM, discussing the principles behind the approach and general applicability.
For the authors, TM is \textit{``the process of analysing a system to look for weaknesses that come from less-desirable design choices''}.
The idea is to consider these deficiencies before developers append features to systems.

Within the TM Manifesto, more recently there has been discussions on establishing secure DevOps in organisations through the use of so-called TM Capability\footnote{Link: \url{https://www.threatmodellingmanifesto.org/capabilities/}.}.
It consists of devising a modular approach to establish a TM program in organisations, by describing measurable and provable practices that translates to actionable objectives.

Another approach that aims to align practical secure DevOps outcomes and make them a reality within software companies is the DevOps Research and Assessment (DORA)\footnote{Link: \url{https://dora.dev/}.}, or simply ``DORA Metrics'', outlined in \cite{forsgren2018accelerate} that investigated metrics for high performance teams.
Secure DevOps metrics were also investigated by Prates et al. (2019)~\cite{prates2019devsecops}, where authors identified the following ones: 1. Defect density, 2. Defect burn rate, 3. Critical risk profiling, 4. Top vulnerability types, 5. Number of adversaries per application, 6. Adversary return rate, 7. Point of risk per device, 8. Number of CD cycles per month, and 9. Number of issues during Red Teaming drills.

\begin{figure*}[!ht]
\centering
\includegraphics[width=.9\linewidth]{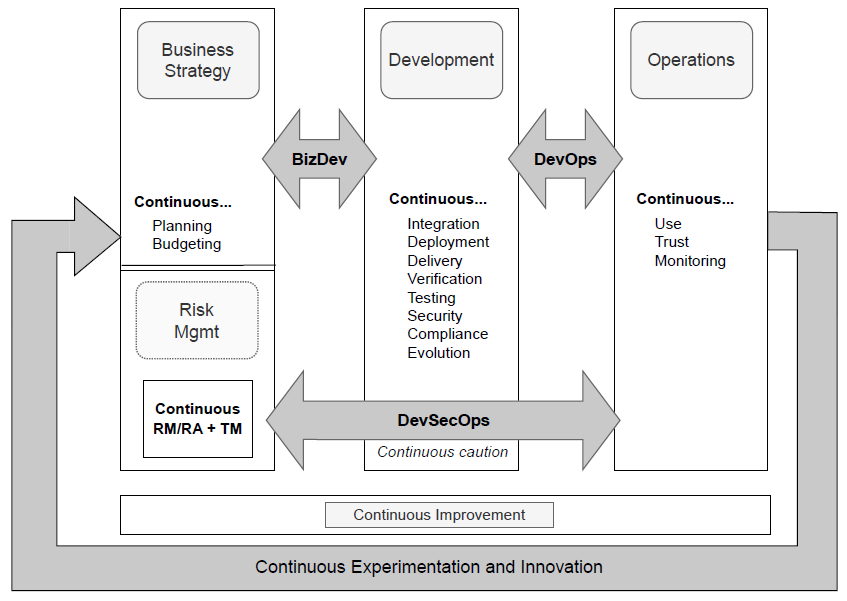}
\caption{Continuous software engineering extended with security concerns and continuous risk -- adapted from Fitzgerald and Stol (2017)~\cite{fitzgerald2017continuous}, by including secure DevOps and continuous RM/RA + TM practices.}
\label{fig:continuous-se-plus-risk}
\end{figure*}

In many organisations there is a mismatch between software development teams and business managers~\cite{fitzgerald2017continuous}.
Going beyond this point we posit that this disconnect is even bigger, as risk analysts are also not participating in the overall quality of products being delivered to customers.
We expand these gaps as discussed in so-called ``continuous software engineering''~\cite{fitzgerald2017continuous} to append continuous RM/RA activities that are deemed crucial for aligning business objectives to product delivery, as depicted in Figure~\ref{fig:continuous-se-plus-risk}.

We argue that the software engineering community is (rightfully) interested in pushing systems towards innovation through experimentation, however, there are additional concerns that must be enforced to ensure customer assurances to their security and privacy.
That is why in the figure there is the ``Continuous caution'' recommendation, as paraphrasing \cite{tarandach2020threat}: \textit{``In our experience, developers live at the speed of \textbf{deployment}. Architects set the speed of \textbf{progress}. Security people run at the speed of their \textbf{caution}''} (emphasis added).

The notion of tackling continuous tasks is crucial in the SDLC, so we clarify secure DevOps with respect to CI/CD activities~\cite{fitzgerald2017continuous}:
\begin{itemize}
   \item \textbf{Continuous Integration:} crucial activity identified in eXtreme Programming that is triggered by series of interconnected phases, ie, compilation, unit, acceptance, or integration tests, code coverage analysis, adherence to code style conventions, and building solutions. It refers to having \textit{releasable software artefacts} and deploying them to some environment (eg, pre-stage, testing, etc.) not necessarily to customers.
   \item \textbf{Continuous Deployment:} consists on releasing software builds to users automatically.
   \item \textbf{Continuous Testing:} integrate testing related activities as close as possible to coding, quickly fixing errors and defects while integrating code bases. The process is automated, and practitioners assign test cases prioritisation to speed up the overall process.
   \item \textbf{Continuous Experimentation:} this is the cornerstone for quickly understanding deficiencies in designs and learning `what works' and `what doesn't work'.
\end{itemize}
Inasmuch as one must focus on principles not on technologies as stressed out by the security community as a plethora of approaches, tools, and mechanisms are in place to help stakeholders, we shall here embark on a technical discussion pointing out how to harden applications altogether.

\subsection{Case study for framing risk in a proof-of-concept: ACME Corp.}
We turn our attention to devising a proof-of-concept with a potential working example outlining our proposed approach.
Note that we have taken this route because organisations will not disclose their risk related activities (nor they should) due to security reasons.
That is the main reason as to why we are devising an exercise on risk that captures the fundamental elements described herein to showcase our RM/RA plus TM approach.

As a \textit{\textbf{disclaimer}}, we represent here a fraction of (what should be) a comprehensive RM/RA approach, to highlight our proposition's benefits to stakeholders.
We are focusing on technologies and innovation as major business activity, and purposely neglecting broader risk relevant tasks (that should be addressed as well) such as natural events, general theft, among other.\\

\begin{figure}[!ht]
\centering
\includegraphics[width=.8\linewidth]{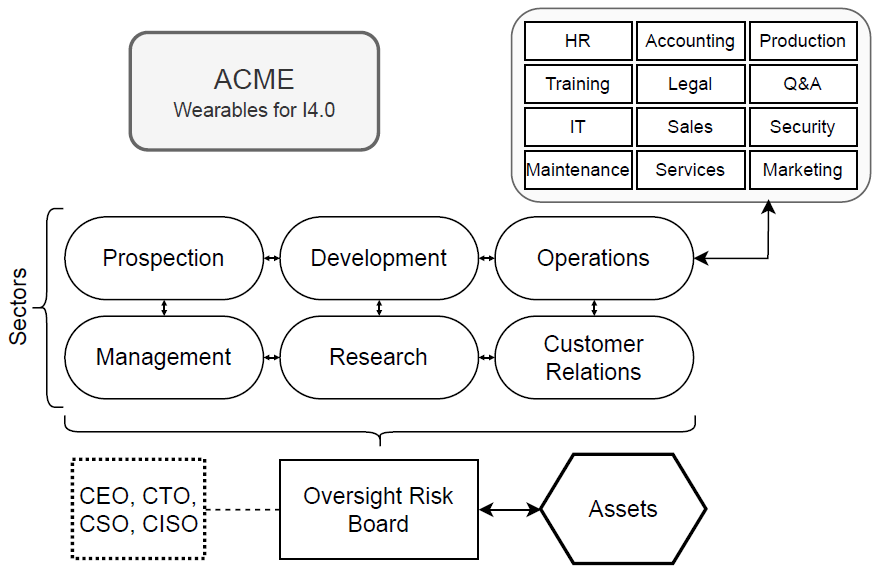}
\caption{Simplified view of ACME Corp. and details for case study outlining sectors and risk related roles.}
\label{fig:case-study-acme}
\end{figure}

\noindent \textbf{Description:} ACME Corp. (a fictitious enterprise) is a for-profit hardware/software organisation with global outreach with offices in London, New York, and Singapore.
The company provides wearable devices (ie, Internet-of-Things -- IoT) tailored to Industry 4.0 (I4.0).
It has 1,000 employees among staff, analysts, developers, and hardware/software (hw/sw) architects.
The company's risk appetite is high; they want to disrupt the wearables market with their products and are willing to take risks.
Recent successful cyber-attacks perpetrated against their virtual infrastructure has pushed management to professionalise their security underpinnings altogether.
After thorough analysis on the attacks they discovered malicious activities performed by competitors posing as insider developers with IT-level credentials.
They have hired additional team of experts across sectors (management, operations, research, and development) and started taking security seriously with continuous assessments and training (just to start).
ACME would like to go one step further and raise awareness on a combined risk approach encompassing different company sectors, where risk analysts inform solutions development through systemic threat modelling.
Figure~\ref{fig:case-study-acme} shows a simplified view of  ACME Corp. detailing sectors and risk related roles.
It depicts some internal entities tailored for tackling risk oversight that ensures and establishes controls for addressing threats and vulnerabilities over assets.

In broader terms, here are the TOP 5 issues they chose to focus and, for each, a mitigation option to reduce the impact of the issue:
\begin{enumerate}
   \item Industrial espionage by foreign and domestic agents.
   \begin{itemize}
       \item Zero-copy policy on premises. No drives on machines. No personal computers. Limited Internet.
   \end{itemize}
   \item Insider threats leading to lost in revenue or illegal sharing of corporate secrets.
   \begin{itemize}
       \item Vetting prospective candidates exhaustively. Enacting whistle-blower policies and rewards. 
   \end{itemize}
   \item Loosing staff members to competition (lacking staff retention mechanisms).
   \begin{itemize}
       \item Offer competitive salaries above the market and participation in company's dividends.
   \end{itemize}
   \item Secure SDLC issues giving rise to vulnerabilities.
   \begin{itemize}
       \item Strong enforcement of Secure Code Review.
   \end{itemize}
   \item Massive dataset generation overwhelming or impairing analysis (ie, data deluge).
   \begin{itemize}
       \item Effective tooling and employing advanced data analysis technique with high-skilled experts.
   \end{itemize}
\end{enumerate}

\noindent \textbf{Risk management overview:}
The management team opted for a hybrid ISO 31000:2018 and NIST 800:30r1 approach to RM/RA.
They want to structure their approach and understand (or at least map) most likely uncertainties as they see might occur.
Starting with RM, they compiled the following ideas:
\begin{itemize}
   \item \textbf{Value}: innovative (cutting-edge) solutions in hardware (embedded computing, wearables/IoT) and software (management systems, web interfaces, and micro-controllers) customised for I4.0. Intellectual property and patents, both granted and prospected.
   \item \textbf{Assets}: researchers, developers, equipment, HW/SW designs, wearable/IoT on-site (being prepared) and in-customers (deployed).
   \item \textbf{Framework}: as leadership, we identify the CEO, CTO, CSO (Chief Scientific Officer), and CISO (Chief Information Security Officer). These roles will enforce the commitment towards risk related issues and propagate the required protections to their business model. Evaluation, Implementation, and Design will be constantly updated every month, in a so-called `Risk Orientation Meeting', with key management roles attending with invited personnel depending on agenda.
   \item \textbf{Principles}: everyone is accountable for each one's respective products (hw or sw). They must account for security measures in place to protect the staff under their responsibility as well as machinery, explaining and teaching on novel approaches to thwart attacks or newest mitigation strategies across the board (involving all sectors). As they want to protect their innovations, managers are to develop and explain controls to retain staff, attract new talent, factor in potential insiders (industrial espionage), and establish measures for malicious detection.
   \item \textbf{Process}: they want to be able to map, understand, and communicate risk in a fast-pace fashion. That is why they used their own patented wearable/IoT technologies deploying it into staff's uniforms and equipment. They want to be able to track and cope with uncertainties and prevent issues from happening based on lessons learned and thorough risk analysis using quantitative and qualitative measures. Enacting continuous Threat Modelling. Controls positioned over the infrastructure capture noticeable events storing them in information systems (SIEM, Firewall, Intrusion Detection System, Issue Tracker, and Event Loggers), complementary to application logging.
\end{itemize}

Additionally, they have established an \textit{Oversight Risk Board} that analyses risks and come up with revisions on the process adapting it to modern times by inspecting latest Advanced Persistent Threats (APT) incursions and motivations particular to this industry.
They respond to upper management, ie, CEO, CTO, CSO, and CISO.
This mechanism will have at most four appointed members (according to expertise) to expedite decisions and achieve faster consensus on every activity involving risk within organisation's sectors.

\noindent \textbf{Risk assessment:} As mentioned, the approach is hybrid: ISO 31000:2018 (\textbf{[A]}) and NIST 800.30 (\textbf{[B]}).
\begin{itemize}
   \item Process [B]: this step will map threats using common techniques and modelling.
   \begin{itemize}
       \item Prepare [A] \& Identification [B]: align business objectives with overall RA process. Understand major sources of risks (as mentioned, ie, espionage, insiders, staff retention, Secure SDLC principles, and data deluge as major risks, other risks and uncertainties could follow).
       \item Conduct [B]: produce a list of security risks with prioritisation to inform decisions. Perform threat and vulnerability analysis, identifying impact, likelihood, and associated uncertainties. Conduct Threat Modelling to map relevant events. Identify crucial systems and security controls in place.
       \item Communicate [A,B]: share results and relevant information from previous phase (conduct) to key stakeholders to guide decision making process.
       \item Maintain [A,B]: keep risk assessment current and updated with latest information about threats and vulnerabilities.
   \end{itemize}
   \item Model [B]: Map out systems, services, throughout the organisation listing threats and vulnerabilities as discovered in previous phases, aligning with business objectives and understanding risk prioritisation over assets.
   \begin{itemize}
       \item Factors [B] \& Analysis [A]: the organisation conducts data-driven analysis over identified threats and vulnerabilities in assets, according to likelihood, impact, severity, and priority. Assess events that could occur to assets.
       \begin{itemize}
           \item Assets [A]: map current assets within the organisation.
           \item Vulnerabilities [A,B]: understand potential weaknesses and places where attacks can take place.
           \item Threats [A,B]: analyse most likely threats with respect to assets.
           \item Event likelihood [A,B]: determine how likely the event is bound to happen.
           \item Impact [A]: address how the event might impact the organisation.
       \end{itemize}
   \end{itemize}
   \item Approach [B]: data-driven (quantitative) with subjective judgement on non-quantifiable (or intangible) data (qualitative). For checking out the SDLC and adherence to Shift-Left Security practices, combined with the use of DORA Metrics to measure teams' performance and response attributes related to mitigation efforts.
   \item Evaluation [A]: make risk related decisions on available information regarding the assessment, comparing with previous analysis, setting additional actions to conform, conduct risk treatment steps to tackle identified risks, and review security controls in place to account for reducing risk to assets. 
\end{itemize}

\noindent \textbf{Threat Modelling:} this step is intertwined with previous one with respect to identifying most likely threats and vulnerabilities aligned to business objectives.
It is crucial to employ modelling that teams with different backgrounds can understand and communicate to others.

Security team strongly suggests adopting TM that provides value to teams and focus on high-level descriptions of systems and possible abuses.
Priority is given to products reaching end-customers on critical environments (eg, healthcare), however, security is viewed as everyone's problem, so issues concerning this element are deployed throughout the portfolio, internally and externally.  
Specialised teams will perform application decomposition analysis over (crucial) HW/SW products to map out specific threats permeating the solutions offered by the organisation.
Following the risk proposition outlined in this work, the organisation wants stakeholders to understand business objectives and align all RA and TM altogether for achieving better results.\\

Teams working on any product (at any stage, conception, design, usability, testing, production, etc.) are encouraged to propose new modelling methods or incorporate different existing TM or concepts into the mix.
These meetings are open for any person at the organisation, and they are also free to invite any stakeholder with specific expertise to help the TM effort.
Teams are expected to tackle CIA+ aspects for any product, system, or service in ACME's portfolio as well as document their processes and update the models altogether.

Specific to this case study, the organisation decided to focus on DFD, Attack Trees, and STRIDE for SDLC related activities (they serve as possible indicators of threat hunting activities, no team member should embrace one technique over the other or push teams to adopt one).
The software tool-chain (versioning, issue ticketing, security systems, event logging) and supporting systems have entries with cybersecurity notes and links to latest incursions and vulnerabilities catalogues, prioritising problems, and determining severity of each noticeable task to tackle next.
As stated, the organisation expects collaborators to keep an open mind about approaches, focusing on potential attack venues instead of specific technologies and modelling alternatives.

\noindent \textbf{Dynamics:}
Enforcing security through risk informed activities throughout organisational sectors:
\begin{itemize}
  \item Alignment to organisational objectives (summary): focused on i) industrial espionage, ii) insider threats, iii) staff retention, iv) Secure SDLC, and v) data deluge (except for iii), activities are highly technological, and security controls must be enacted accordingly, observing CIA+ objectives).
  \item Map current attack surface effectively by means of information systems combined with secure logging.
  \item Controlling data within organisation: authentication and authorisation mechanisms in place to understand access and perform auditing and forensics if ever required.
  \item Shift-Left Security by involving developers and architects in secure programming and design practices from the onset of activities. Match User Stories to Security User Stories that map CIA underpinnings directly into the process.
  \begin{itemize}
      \item Test coverage: it continuously tests how much of the code is being tested.
  \end{itemize}
  \item Secure Code Review: experts review all HW/SW code and allow only secure tested code and vouched API to enter the repository.
  \item Observability and secure logging: identifying and alerting malicious activities (as perpetrated by insiders).
  \item Data analytics: de-duplication, processing, treating, analysing, converting, and analysing data, making sure data in transit and in rest are confidential.
  \item Quick detection, alerting, and mitigation: allocating response teams if any threat appears and establishing actions to minimise damage and secure the operation.
\end{itemize}

Upper management understands the importance of risk within the organisation and identifies opportunities for continuously improving the overall RM/RA process integrated with TM.
Risk analysts are expected to engage across sectors for communicating risks across the board.
Security officers engage with risk analysts to align objectives altogether.
SDLC developers and architects engage with security officers and risk analysts for improved Shift-Left Security practice.


\section{Recommendations and guidance}\label{s:recommendations}
Because of its broad scope, RM indicates risk strategy on an organisational level, outlining leadership, roles, and responsibilities.
Acting more locally, RA informs tactics for threats.
TM tackles working with devising ways of abusing systems and technologies.

\subsection{From potential features to concrete features in production}
\begin{figure*}[!ht]
\centering
\includegraphics[width=.9\linewidth]{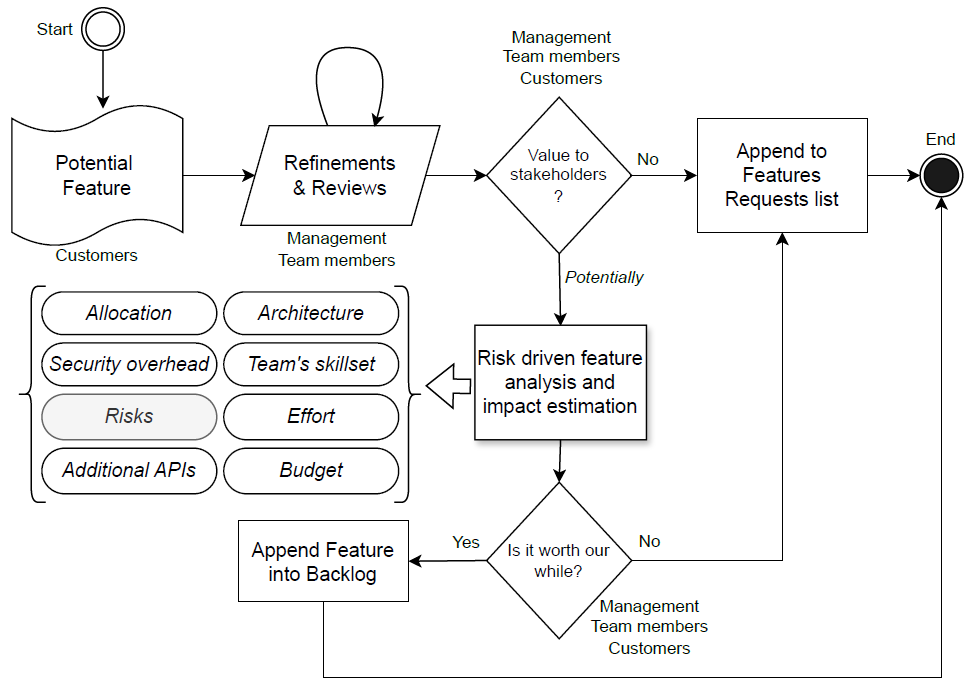}
\caption{Risk-informed process for inserting potential features into the Backlog in secure DevOps.}
\label{fig:journey}
\end{figure*}

Before they become actual features executing in production, raw ideas and wished for functionalities exist only in customer's wish lists or resting in (non-prioritised) backlogs.
They must add value to the software solution, have an associated cost and require equally costly resources to be allocated.
Figure~\ref{fig:journey} showcases this process and outlines the major concerns and decisions required by stakeholders when deciding whether these undefined ideas should become full-fledged User Stories (features) and enter the Product Backlog.

And because secure DevOps and Shift-Left Security is supposedly enforced into the process, additional concerns (potentially leading to delays) are also in play.
Under such context, team leaders and managers must thoroughly study the feature's impact on their operation before any development takes place.
We outline next the journey of such raw ideas until they become operational in production to understand the required effort and budget concerns to factor in.

Before we take on this task, let's devise the following assumptions:
\begin{itemize}
   \item As stressed here, teams must understand the RM/RA context to guide their security-based decisions altogether. Relating RA and TM within secure DevOps entails understanding architectural details for software solutions and mapping inherent risks associated with implementation and operations embedding it all with security requirements.
   \item Secure DevOps presumes the productive existence of continuous feed-back loops among stakeholders, ie, customers, team members, security officers, domain experts, and managers.
   \item Project adopted an incremental approach with frequent releases and aiming stakeholder feedback throughout phases.
   \item The team has decided to use, as software development process, Agile approaches, perhaps even considering Scrum, XP or Kanban (to mention some). Under this, decision teams will work with a list of User Stories (appending them into a Product Backlog), breaking them, adding them to Sprint Backlog, and assigning required effort.
   \item Listing vulnerability catalogues to track, observe, and align with the SDLC.
   \item Use of a Versioning system (Git based) employing Trunk Based Development where main branch is the latest software version.
   \item Secure code review in place: security experts review code before they are committed into the versioning system.
   \item Shift-Left Security and early TM: security is under consideration since requirements/plan/design phases and run throughout the SDLC, especially in early phases.
\end{itemize}

\begin{figure*}[!ht]
\centering
\includegraphics[width=.95\linewidth]{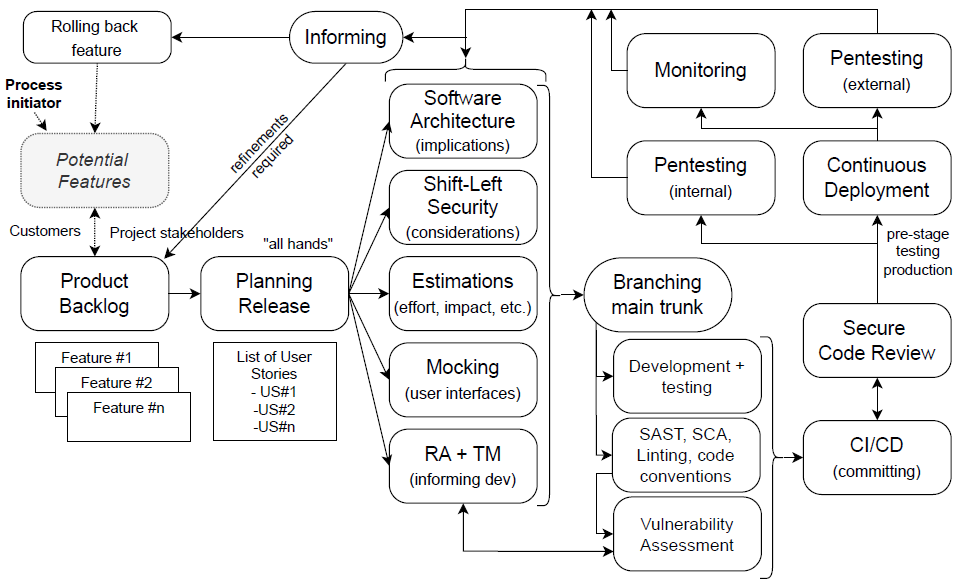}
\caption{Overall secure DevOps process intertwined with SDLC, RA, TM, and Shift-Left Security.}
\label{fig:devsecops-process}
\end{figure*}

Figure~\ref{fig:devsecops-process} illustrates the overall process encompassing secure DevOps with respect to SDLC and addressing RA, TM, and Shift-Left Security altogether.
Next stage consists of moving forwards into a Secure SDLC approach intermixed with secure DevOps cycle, for instance:
\begin{itemize}
   \item \textbf{Requirements}: identifying, mapping, refining, and reasoning about implementation effort and usefulness (value).
   \item \textbf{Plan and Design}: understanding which modules will be required for change, which teams they might impact for extra testing and integration.
   \item \textbf{Risk Assessment and Threat modelling on early designs}: as mentioned, RA and TM tasks can kick-start early, with drafts of architectural documentation and application mocks (user interfaces) for devising DFD or other models that inform better design decisions to stakeholders.
   \item \textbf{Implementation} (branching from versioning system) and \textbf{integration} with (other) team's source code.
   \item \textbf{Testing}: creating and updating Unit Tests (ie, automated validation), involving team commitment for devising and reviewing through additional implementation.
   \begin{itemize}
      \item \textbf{Application Security Testing (AST)}: focus on SAST, linting, adherence to code conventions (as set by the team), and Fuzzy Testing.
   \end{itemize}
   \item \textbf{Software Composition Analysis (SCA)}: checking external API/libraries, versions, and licensing, as well as checking for vulnerabilities.
   \item \textbf{Automated Threat modelling}: using software artefacts as available in projects to compose TM amenable for analysis and guiding software teams developing high-quality code and adherence to security.
   \item \textbf{Secure Code Review}: security experts comment on code and uphold constraints, suggest changes, and check whether Unit Tests were updated to meet feature's objectives, and compliance to regulation and standards.
   \item \textbf{Integration}: merging code, solving conflicts, re-running tests, integration testing.
   \item \textbf{Delivery}: commit to main branch (deliverable version of feature or set of features)
   \item \textbf{Internal pentesting}: mind that this is a biased approach as one is familiar with the system and might only test what is working, not a comprehensive test suite aiming at finding active vulnerabilities.
   \item \textbf{Deployment}: feature set is pushed into test, pre-production, staging or production environments.
   \item \textbf{DAST}: executing software in a specific environment to track potential sources of vulnerabilities. As mentioned, this process is highly time-consuming and it involves the allocation of costly resources, where the gains and insights are sometimes disputed by teams as this process could be replaced by SAST and testing. 
   \item \textbf{External pentesting}: conducted by third-parties (unbiased) where the team is given a report for addressing issues found in the process.
   \item \textbf{Monitoring}: secure DevOps Metrics (DORA based), tracking features, network usage, application logs, system statistics
   \item \textbf{Maintenance and evolution}: overseeing system under execution for quality improvements.
\end{itemize}

\subsection{Approaches: Start with Security, Shift-Left Security, or Start-Left}
The Federal Trade Commission (FTC) in the US devised a White Paper in 2015~\cite{federal2015start} inviting organisations to ``Start with Security'', listing a series of guidance such as 1. Control access to data, 2. Enforce secure passwords and authentication, 3. Store sensitive information securely (in rest and in transit), 4. Segment your network, 5. Grant secure remote access to your network, 6. Apply sound security practices, 7. Make sure your service providers meet your secure standards, 8. Keep security current and address vulnerabilities, and 9. Secure paper, physical media and devices.
This document is a blueprint for addressing security concerns early in systems design, discussing physical security elements and cyber related concerns.

The practice of Shift-Left Security (SLS) is inherently collaborative~\cite{rajapakse2022challenges,pargaonkar2020future}, encompassing a range of stakeholders concerned with the security posture of the organisation.
The work by Lombardi and Fanton (2023)~\cite{lombardi2023devops} discussed the need for effective SLS in secure DevOps.
The authors argue that their approach dubbed CyberDevOps\footnote{Patrick Debois has discussed the ``Shades of DevOps Roles'' in a post available at \url{https://www.jedi.be/blog/2022/02/11/shades-of-devops-roles/} (2022), where he outlines how hyped and charged the term \textit{DevOps} has become.} is more effective than straightforward secure DevOps because they incorporated SCA to deal with nondeterministic environments and tackled vulnerability assessment and compliance by adding another pipeline step in the process.

\begin{figure*}[!ht]
\centering
\includegraphics[width=.9\linewidth]{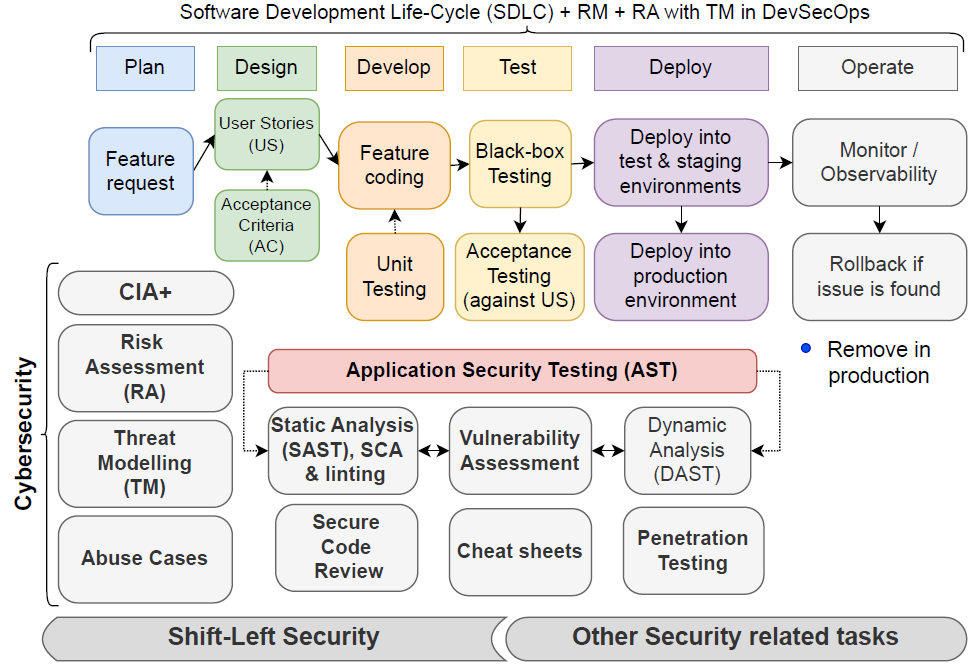}
\caption{SDLC, Shift-Left Security and positioning within a secure DevOps context.}
\label{fig:shift-left-dev-sec-ops}
\end{figure*}

The work by Jimenez et al. (2019)~\cite{jimenez2019devops} discussed SLS in an industrial case study.
They discuss a framework for automating a deployment setting for distributed software systems and components.
For instance, Fisher (2023)~\cite{fisher2023application} advocates employing SLS for preventing costly maintenance and code rewrites that may or may not introduce additional vulnerabilities into applications.
The author comments on the benefits of this approach and cite examples for achieving success, for instance, making good choices on the right architecture and other design decisions early whilst taking security into consideration.
It includes using institutions trusted by industry for instance, OWASP, that foments training and guidance when hardening applications.
Other interesting ways to consider include protecting data flows and database technologies, eliciting requirements on encryption strategy with visible documentation throughout stakeholders, among other suggestions.
Figure~\ref{fig:shift-left-dev-sec-ops} shows the SDLC coordinated with Shift-Left Security approaches plus secure DevOps cycle, AST, and tooling.

\subsection{Additional quality attributes required by software projects}
One could posit that other software quality properties could also be shift-left in design and planning phases, for instance, usability, maintainability, scalability, modularity, performance, and more recently, authors recognised energy efficiency as an attribute~\cite{bass2021software}.
The issue in software development is that teams are almost never sure which dimension is more important than the other, and it requires experience to ``get things right the first time''.
The decision on which one to focus could impact project's speed and productivity altogether, so they must carefully analysed and aligned with other project stakeholders.
These properties are trends and \textit{buzzwords} that should be taken seriously by software developers, ie, the case of a project's objective to be robust, reliable, resilient, adaptable, survivable, or sustainable.
Serious projects address selected quality attributes with actionable tasks that effectively ensures that it addresses the issues and that it sustains it with measurements and metrics.

Although this is intrinsically related to software architecture, it has ramifications in secure DevOps~\cite{bolscher2019designing,shahin2016intersection}.
The work by Alnafessah et al. (2021)~\cite{alnafessah2021quality} discussed quality-aware DevOps, identifying research challenges to improve architectural design, modelling and Infrastructure as Code, CI/CD, testing and verification, and runtime management.
The latter also recognises the intermixing with emerging trends of Artificial Intelligence (AI) for DevOps, approaches they expect to dominate research in the years to come.

When teams are modelling software they must work alongside with software architects that share the product's design and allow security issues to be raised by security officers.
With this documentation they could kick-start TM right away and inform development teams of potential vulnerabilities and points to focus on.
Usual culprits are input handling, communication, data in rest and in transit, session management, and cryptography, to mention some concerns.
These decisions and guidance are related to (future) maintainability and secure code review, where more experienced programmers advise on `dirty tricks' that could lead to unintended exposure, data leaks, or weaknesses.

Modern software solutions are inherently complex, dependent on multiple auxiliary libraries and Application Programming Interfaces (API) to allow the seamlessly provision of the solution in testing and production environments.
These quality related notions circle back to addressing software complexity, scalability (at the right moment), premature refactoring for performance, and multiple API integration in projects (software inter-dependencies).
In terms of scalability, teams must estimate application usage and adjust the architecture to meet this demand.
There is a need to break solutions down into more manageable parts by means of so-called application decomposition, helping out establishing adequate modularity with implications on TM and testing.
Beck's book \cite{beck2023tidy} thoroughly discussed coupling/cohesion, explaining how software design functions and how to address making large changes in software through small incremental steps.
With respect to risk and TM, PASTA~\cite{ucedavelez2015risk} has a step devoted to application decomposition in its threat modelling process.
Finally, software evolution is related to maintainability and addressing technical debt~\cite{brown2010managing,yli2016software,behutiye2017analyzing}, ie, a situation where software developers prioritise speed to push changes that are easy (low hanging fruit) instead of really fixing the issue, that invariably takes more time and resources.

\subsection{Technological guidance and tooling}

\begin{figure}[!ht]
\centering
\includegraphics[width=.9\linewidth]{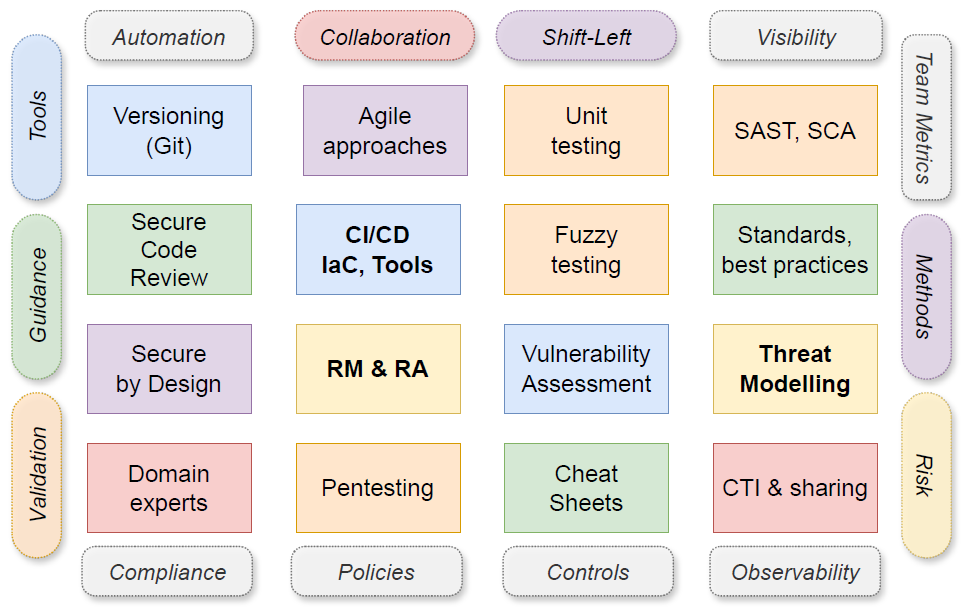}
\caption{Secure DevOps characteristics \& practices in SDLC mixed with RM/RA.}
\label{fig:devsecops-practices}
\end{figure}

Figure~\ref{fig:devsecops-practices} shows a non-exhaustive list of technologies and practices enabling secure DevOps that improves productivity.
It focuses on the ones that are deemed useful by software teams over the host of results over the years as commented out earlier.

Embracing secure DevOps in software organisations demand a cultural change that encompasses different sectors and departments, ie, risk analysts, developers, domain experts, and customers involved in decisions.
The risk faced by many software organisations is to have unmaintainable bloatware full of unknown vulnerabilities that are susceptible to costly cyber-attacks.
To make things worse, if the team does not fully embark in the security oriented process, overtly testing (white,gray,black-boxes) and updating both tests and (threat, data) models, the secure DevOps initiative might fail in the medium to long term.
That is why it is important to set up expectations from the beginning, trying to remove any (cultural, communication, ego, etc.) barriers within development teams, aiming a successful operation.
Additionally, it is expected to align these secure DevOps practices with other stakeholders, namely upper management (CEO, CSO, CTO, CISO\footnote{Respectively Chief Executive Officer, Chief Scientific Officer, Chief Technology Officer, and Chief Information Security Officer.}) with risk analysts, domain experts, development/testing teams, and customers (inadequate involvement is detrimental to project success).
Note that secure DevOps imposes extra activities upon stakeholders, for instance, setting up tool-chains (choosing most appropriate tools), and configurations, on top of hiring security experts that are able to communicate ideas with members.
The project documents should encompass not only architectural and design notes and observations but also RM/RA approaches and methodologies.
Finally, note the intentional omission of DAST in the figure.
This technique is seen as a time-consuming task and does not yield significant outcomes for secure DevOps practitioners~\cite{rajapakse2022challenges}.

In the secure DevOps community there are recommendations for using tools to automate security readily incorporated into the CI/CD pipeline.
However, teams must exercise caution for understanding the need for frequently attesting its `freshness', ie, whether or not they remain being useful in the project.
Additionally, it is recommended to use community vouched mechanisms that adds value to the project, and one example is provided by OWASP's cheat sheet series, that addresses multiple concerns for incorporating security in varied scope projects.

\section{Conclusion and roadmap for integrating risk assessment in secure DevOps}\label{s:conc}
Broader risk informed approaches are valuable for understanding and mapping most-likely adversaries and organisational focus on value creation.
The  motivation for the discussion presented here is due to considering security as an afterthought and simply assume projects seriously consider it through and throughout.
The seminal work by Fitzgerald and Stol (2017)~\cite{fitzgerald2017continuous} commenting on continuous-* in software engineering very shyly mentions security, trust, and privacy, focusing instead on development and totally disregarding risk and threat modelling altogether.
We stress the need to combine risk approaches and align it with business strategy and objectives in SDLC where secure DevOps and CI/CD must be thoroughly considered in a RA and TM context.

Unfortunately, for many organisations, security considerations are not the focus of the approach, and many times are just an afterthought.
It is often taken serious whenever subject to actual attacks with negative repercussions.
The usual proportion within companies is 90:9:1, meaning a ratio of 90 developers, 9 IT/operation, and 1 security officer or someone interested in cybersecurity.

Security personnel should not consider RM and RA independently from TM: these are complementary approaches aimed at ensuring high end-product quality in systems and services to customers.
As discussed throughout this contribution, the vast body of knowledge of cybersecurity invites practitioners to focus and not waste time on adopting unproven approaches.
That is the main reason why it offers sets of recommendations, good practices, lessons learned, and guidance so security experts approach risks and threats over assets by aligning the investigation with each one's organisational objectives.
Secure DevOps is suited for Agile as in Waterfall there is the egregious approach to finish projects and only then start to think about security (``let's do security now'' mindset).
This is why this method is detrimental to modern SDLC as teams append security concerns too late, a practice leading to additional costs.

Experienced managers know that adding security into any software project is costly and it slows down team velocity.
This is compounded by the team choosing to adopt a complex technology stack that adds up to risk as additional software gets associated, sometimes with minimal security testing.
The key is to train your team to quickly adapt to the new secure DevOps reality in a seamless fashion, making them understand the need for establishing the secure oriented culture and the projected future gains of this choosing.

The work by Grigorieva et al. (2024)~\cite{grigorieva2024development} discussed five pillars of secure DevOps: 1. CI/CD, 2. Automation and vulnerability scanning, 3. Secure coding practices, 4. Container security, and 5. Shift-Left Security whereas Alnafessah et al. (2021)~\cite{alnafessah2021quality} complement the list with infrastructure as code, verification, and runtime management.
These authors point out the challenges for it to become a reality in organisations, for instance, cultural shift, tool integration, compliance alignment with regulatory standards, skills gap and professional shortage, and speed vs security balance, ie, security trade-offs in projects.
We next complement this discussion with extra points to observe:
\begin{itemize}
  \item Training staff for security: making stakeholders understand that \textit{``security is everyone's problem''}~\cite{woods2021continuous} and also identifying potential champions to push forward this `vision' across the project.
  \item Avoid ``security theatre'': this notion is for when organisations state that they care deeply about security, however they put few measures in place or allocate small budget to effectively tackle security issues encompassing their attack surface and services portfolio.
  \item Building team trust: establishing a relationship among team members, even if virtually.
  \item Mixing cybersecurity professionals in heterogeneous teams (coders, testers, managers, domain experts): they have a broad understanding of security issues arising in software projects and are able to consider it from the onset of the design and implementation phase.
    \begin{itemize}
        \item As described in the literature~\cite{fisher2023application}, usual development teams are formed with assigning one security champion to absorb all security related work (sometimes it is not his/her expertise) -- when vulnerabilities are found (by any method), they get overwhelmed.
    \end{itemize}
  \item Mapping potential threat actors according to your organisation: which adversaries are most likely to target your setting? What are the common venues employed by these cyber-attackers in your domain? How to deal/track insider threats?
  \item Frequent update of RA and TM: it aims to quickly absorb changes in infrastructure, architecture, and novel attack-venues as sophistication increases.
  \item Understand what you are building: Domain Driven Design (DDD)~\cite{evans2004domain,vernon2013implementing,khononov2021learning}, bringing stakeholders to the development, shared ubiquitous language and problem understanding.
     \begin{itemize}
         \item Organically considering Secure by Design (SbD)~\cite{deogun2019secure} thinking within the development team, exploring more secure alternatives that could result in less maintenance effort in future changes (related to secure code review).
     \end{itemize}
  \item Established secure code review practice: cybersecurity experts review code in search for potential vulnerabilities, future maintenance issues, and `code smells' that could turn into serious defects or costly refactoring. 
  \item Managerial, ie, related to project management:
      \begin{itemize}
      \item Avoid micromanaging software teams: adopt full-Agile approach where the team is aware of its responsibilities without the figure of a `Scrum Master' or `Product Owner' that sometimes delay team's speed\footnote{This is \textit{highly contentious} as the Agile community has been discussing means for effectively embracing a \textit{true Agile} instance and how the Scrum framework and similar interventions in teams are detrimental to software teams. It is out of the scope of this work to discuss these issues.}.
      \item Effective meetings: pre-defining action items and assigning responsibilities and points for discussion; maximum recommended time spent is 10-15 min per meeting.
          \begin{itemize}
              \item Developers need (large) chunks of focus intensive time to be productive.
          \end{itemize}
      \end{itemize}
  \item Core Software Engineering:
      \begin{itemize}
          \item Understanding the effects of cohesion and coupling: reasoning how to adequately decompose systems into parts (modules), and how much interaction they should sustain is critical for system's designs that scale (ie, from monoliths to microservices architectures).
          \item Code smells: there is a need to identify good programming practices and differentiate them from bad design choices very early in the process, in combination with code review from more experienced developers and security experts, as mentioned. 
      \end{itemize}
  \item Implementation choices (technical):
      \begin{itemize}
        \item Trunk-based development: the main branch has the stable version that has passed secure code review successfully.
        \item Effective, validated refactoring (ie, supported by unit testing): making code improvements without changing functionality has been happening for a long time, the difference is doing this without introducing defects into the repository.
        \item Network segmentation as a security feature: it isolates services or features altogether, increasing protections against users.
      \end{itemize}
  \item Speed/Security trade-offs: consider the renowned issue of asymmetry in cybersecurity research~\cite{xu2021seeking}, where attackers only need to identify one single vulnerability to exploit a resource whilst defenders must seek to prevent or block all vulnerabilities. Now compound with the fact that there is an abysmal difference in terms of the number of a. programmers, b. IT professionals (operations), and c. security experts.
    \begin{itemize}
       \item Teams and management should strike a balancing act in their projects when coping with these shortcomings, aiming at operational sustainability to keep the business thriving.
    \end{itemize}
   \item Dealing with teams: note that \textit{clever} developers will try to by-pass your guidance and constraints, for various reasons, ie, perceived delay/speed, provoke security officers, or even seeing the issue as a technical challenge (or just showing off).
  \item Observability: inspecting logs, traces, and measurements, tools or human-in-the-loop counterparts can seamlessly understand attack progression and devise mechanisms to thwart malicious incursions before they spill over other systems and services~\cite{majors2022observability}.
      \begin{itemize}
          \item Establishing multiple data sources and then applying automated de-duplication tools (to remove duplicated entries) could help identifying Advanced Persistent Threats (APT) Living Off The Land (LOTL) and insider threats within organisations.
      \end{itemize}
\end{itemize}

Recent advances in AI providing coding assistants and thorough lessons learned and analysis from other projects may change the DevOps landscape altogether.
At the time of writing, projects wishing to adopt secure DevOps require humans-in-the-loop for determining success, refine prompts, and review automated responses for accuracy/veracity.

It is a fact that security concerns impose additional guard-rails to any development process.
If this process could be cautiously automated to take into account security requirements that managers and team members trust, that will have profound effects on productivity.
Security officers want systems and services to be secure (sacrificing velocity) in a context where developers want flexibility when developing solutions.
Ideally, project team members strike a balance that allows everyone to be productive and as secure as possible.

As future work we aim to put the considerations discussed herein into a practical software project and evaluate its implication in real-world settings.
We firmly believe that secure DevOps practices integrated with risk assessment matures in the community and there is a broader understanding for intermixing these concepts altogether.

\bibliographystyle{unsrt}  
\bibliography{bibliography}  

\end{document}